\documentclass[aps,prl,twocolumn,showpacs,groupedaddress]{revtex4}  
\usepackage{graphicx}  
\usepackage{dcolumn}   
\usepackage{bm}       
\usepackage{amssymb}   
\usepackage[section]{placeins}
\usepackage{capt-of}
\usepackage{color}
\usepackage{amsmath}
\usepackage{mhchem}

\begin{document}
	
	\widetext
	\title{High-energy polarized electron beams from the ionization of isolated spin polarized hydrogen atoms. }

\author{Dimitris Sofikitis$^{1}$}
\email{sofdim@uoi.gr}
\author{Lars Reichwein$^{2,3}$}
\author{Marios G. Stamatakis$^{4}$}
\author{Christos Zois$^{1}$}
\author{Dimitrios G. Papazoglou$^{5,6}$}
\author{Samuel Cohen$^{1}$}
\author{Markus Büscher$^{3,7}$}
\author{Alexander Pukhov$^2$}
\author{T. Peter Rakitzis$^{5,8}$}

\affiliation{$^1$Department of Physics, Atomic and Molecular Physics Laboratory, University of Ioannina, University Campus, Ioannina, GR-45110, Greece}
\affiliation{$^2$Institut f\"{u}r Theoretische Physik I, Heinrich-Heine-Universit\"{a}t D\"{u}sseldorf, 40225 D\"{u}sseldorf, Germany}
\affiliation{$^3$Peter Gr\"{u}nberg Institut (PGI-6), Forschungszentrum J\"{u}lich, 52425 J\"{u}lich, Germany}
\affiliation{$^4$Department of Mathematics, University Campus, Ioannina, GR-45110, Greece}
\affiliation{$^5$Institute of Electronic Structure and Lasers, Foundation for Research and Technology-Hellas, 71110 Heraklion-Crete, Greece.}
\affiliation{$^6$Materials Science and Engineering Department, University of Crete, GR 70013, Heraklion, Greece}
\affiliation{$^7$Institut f\"ur Laser- und Plasmaphysik, Heinrich-Heine-Universit\"{a}t D\"{u}sseldorf, 40225 D\"{u}sseldorf, Germany}
\affiliation{$^8$Department of Physics, University of Crete, 70013 Heraklion-Crete, Greece.}

\begin{abstract}
			
We propose a laser-based method for the preparation of high-energy polarized electrons, from the ionization of isolated spin-polarized hydrogen (SPH) atoms. The SPH atoms are prepared from the photodissociation of HCl, using two consecutive UV pulses of ps duration. By appropriately timing and focusing the pulses, we can spatially separate the highly polarized SPH from other unwanted photoproducts, which then act as the target for the acceleration lasers. We show how elastic collisions define number density $n$ and polarization P regimes ($10^{16}\leq$ $n$ $\leq 10^{18}$ cm$^{-3}$, 0.99 $\geq$ P $\geq$ 0.40) for the pre-polarized targets, and use particle-in-cell simulations to demonstrate the method's feasibility. 

\end{abstract}	

	\date{\today}
	\maketitle

Polarized electron and positron beams are powerful experimental tools, used in a diverse set of disciplines, ranging from studies of atomic and molecular structure \cite{kessler1990,gay2009} and material science \cite{PlasmaTrapBazed,SolidStateRevModPhys}, to nuclear and high energy physics \cite{sun2022production}, where electrons accelerated to relativistic energies can be used to test new physics beyond the standard model \cite{ILC2013,androic2019precision,Neutron2013}.
Producing intense beams of highly polarized, high-energy electrons can be done using conventional acceleration methods involving accumulation in storage rings \cite{sokolov1986radiation,ELSA}, as well as emerging methods involving filtering of polarized electrons \cite{SpinFilter,Splitter}. Alternatively, high-energy polarized electrons can be produced using polarized photocathodes \cite{Photocathodes,ElectronGun}, or through laser ionization of noble gases \cite{IonizationNobleGases,NiePRL2021}. However, the maximal electric current for these methods is limited below 0.1 A.

\begin{figure}[hb]
	\centering
	\includegraphics[width=0.45\textwidth]{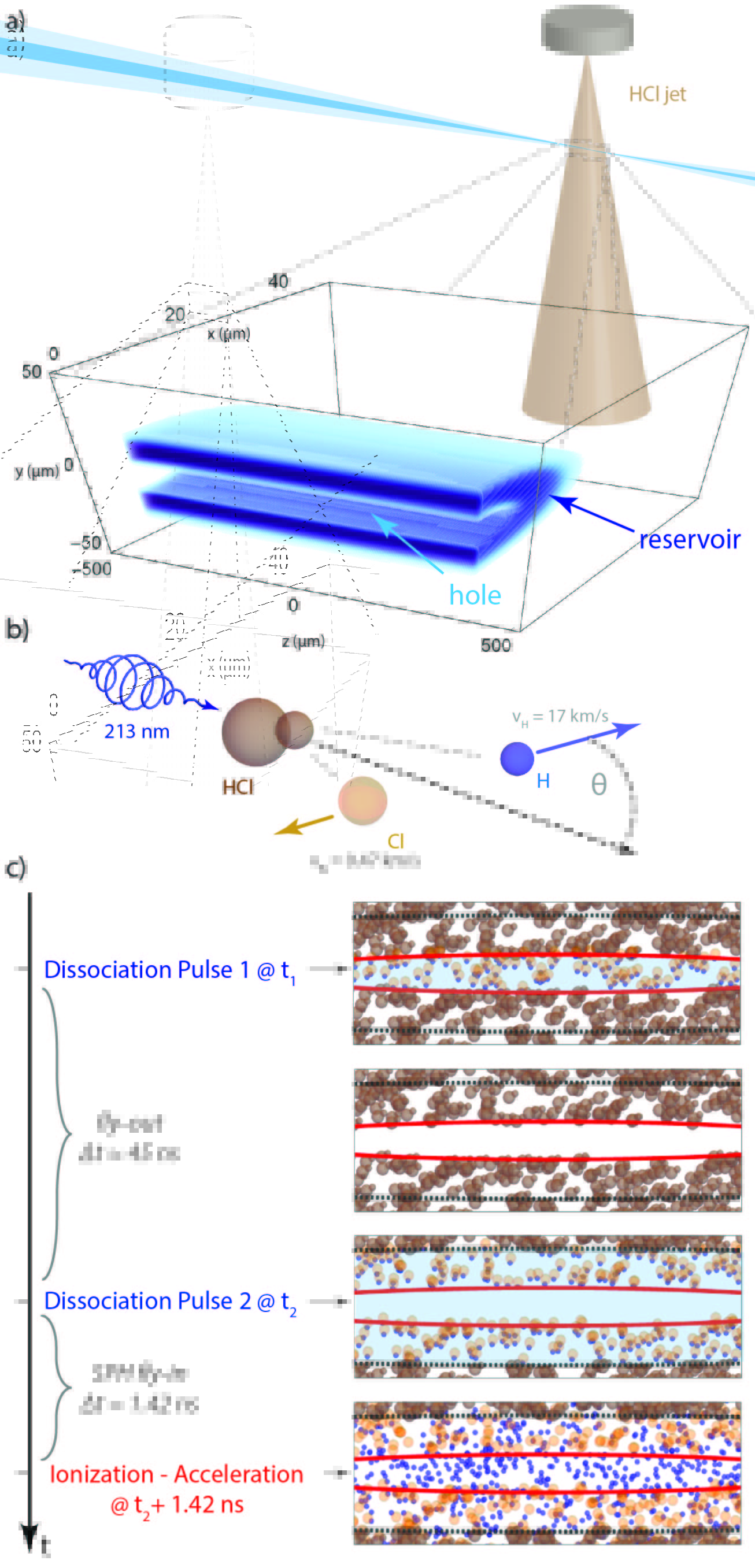}
	\caption{a) Experimental setup, showing the gas jet and laser beams. b) Velocities associated with the dissociation process. c) The  stages of the experiment. (1) complete photodissociation only in area of hole; (2) hole empties of atoms; (3) complete photodissociation in reservoir around hole; (4) hole fills with SPH.}
	\label{FigureSetUp}
\end{figure}

Laser acceleration of electrons \cite{tajima1979laser}, combined with recent laser methods for the preparation of high-density SPH \cite{PtrScienceSPH,Qbeats,spiliotisLSA}, open a new potentiality for orders-of-magnitude higher electric currents: that of accelerating electrons resulting from the ionization of pre-polarized targets. Following this, a variety of recent proposals predict the production of high-energy spin-polarized electron beams utilizing such targets \cite{DualWake2024,KilloAmpere,CPI2023,wu2019polarized}, predicting up to kA electric currents \cite{KilloAmpere}. 

The direct use, however, of such SPH atoms as targets for electron acceleration is limited by several factors. The presence of the halide atoms means that their valence electrons (which are only weakly polarized by the photodissociation) and inner-shell electrons (which are unpolarized) are also liberated and accelerated, lowering the total polarization of the accelerated electrons to very low values. It could be possible to ionize the halide atom and subsequently remove the ions using electric fields~\cite{wu2019polarized}, however, this is quite challenging to achieve within the timescales required for laser acceleration. Finally, dissociation of $H_2$ molecules at wavelengths below 100 nm, which would not suffer any halide atom presence, has not been experimentally tested and the percentage of direct molecular ionization has not been evaluated \cite{KilloAmpere}.

Additionally, the SPH polarization depends on the parent molecule bond orientation, resulting in a cos$^2\theta$ spatial distribution of the polarization, a fact that sets an upper limit for the free-space value for polarization of 40\%. Bond orientation, which can lift this limitation, can be achieved using a strong IR pulse. However, apart from the complication or using an extra pulse of different wavelength (MIR), bond alignment cannot be 100\% successful for moderate IR pulse intensities needed to avoid unwanted multi-photon or even field ionization effects. Finally, the hyperfine structure of hydrogen atoms causes the polarization to oscillate from the electron to the proton and backwards, with a period of 0.7 ns, meaning that any manipulation aimed to removing the unwanted halide atoms or to achieving bond orientation, has to be synchronized with this oscillation to avoid further reduction of electron polarization. Proposed solutions lead to experimental complications, ultimately limiting the method’s feasibility. 

Here, we propose a simple and intuitive method which circumvents all the limitations mentioned above. The method takes advantage the kinematics of the dissociation process, the angular distribution of the polarization of the atomic fragments and the shape of the dissociation laser beams, to produce pure targets of highly polarized SPH, without the presence of the unwanted halide partners.

In the first step, we illuminate an HCl sample using a focused UV laser to dissociate all molecules within the laser focus and along the Rayleigh range of the laser beam. The Cl and H atomic fragments will acquire large velocities, that will cause them to exit this volume within a few tens of ns, creating a volume devoid of atoms or molecules, hereafter referred to as 'hole'. In the second step, a second dissociation pulse of larger spatial dimensions (but of same wavelength and pulse duration as the first), dissociates molecules in a larger volume around the hole, hereafter called the 'reservoir'. SPH atoms will now fly from the reservoir to the hole volume (much faster than the heavier Cl atoms), where they can be ionized to produce accelerated electrons by a an acceleration pulse in the third step. The narrow velocity distributions of the fragments and the available dissociation laser beam geometries allow tailoring the hole and reservoir geometries, so that, in the third step, only hydrogen atoms of high polarization are contained inside the hole.

We consider Dissociation Pulses 1 (DP1) and 2 (DP2), of the same wavelength and duration (around 200 nm and 100 ps for example), but with different spot size, intensity and synchronization. The pulses are directed towards a molecular beam containing HCl molecules (Fig.\ref{FigureSetUp}). The molecular beam moves with a velocity around 1000 m/s  perpendicular to the direction of propagation of the dissociation pulses; for simplicity, we use the reference frame of the moving molecular beam. 

At t = 0, the pulse DP1 will dissociate all HBr molecules in the volume of the hole, which is shaped as a prolate ellipsoid, with semi-axes a, b ($\approx$ 10 $\mu$m) and c, which are parallel to x, y, and z axes, respectively. We choose the Rayleigh range to be larger than the width of the molecular beam; this way, the hole is truncated, i.e. the front and back ends of the ellipsoid defining the hole are cut, and consequently no HCl molecules are in front or after the area of the target. Dissociating HCl at $\lambda$ = 200 nm results in H atoms with speed $v_H$ $\approx$ 17 km/s and Cl atoms with speed $v_{Cl}$ $\approx$ 0.47 km/s (Fig.\ref{FigureSetUp}b). Thus, the H atoms will leave the volume of the hole in a few ps, and subsequently, the Cl atoms will also exit in few tens of ns, leaving the volume of the hole devoid of atoms and molecules. After $\approx$ 45 ns, DP2 can be fired, which will produce fast SPH atoms in the reservoir, which will then rapidly fill the hole after $\approx$ 1.42 ns (twice the hyperfine beating time). 

\begin{figure}[htbp!]
	\centering
	\includegraphics*[width=0.45\textwidth]{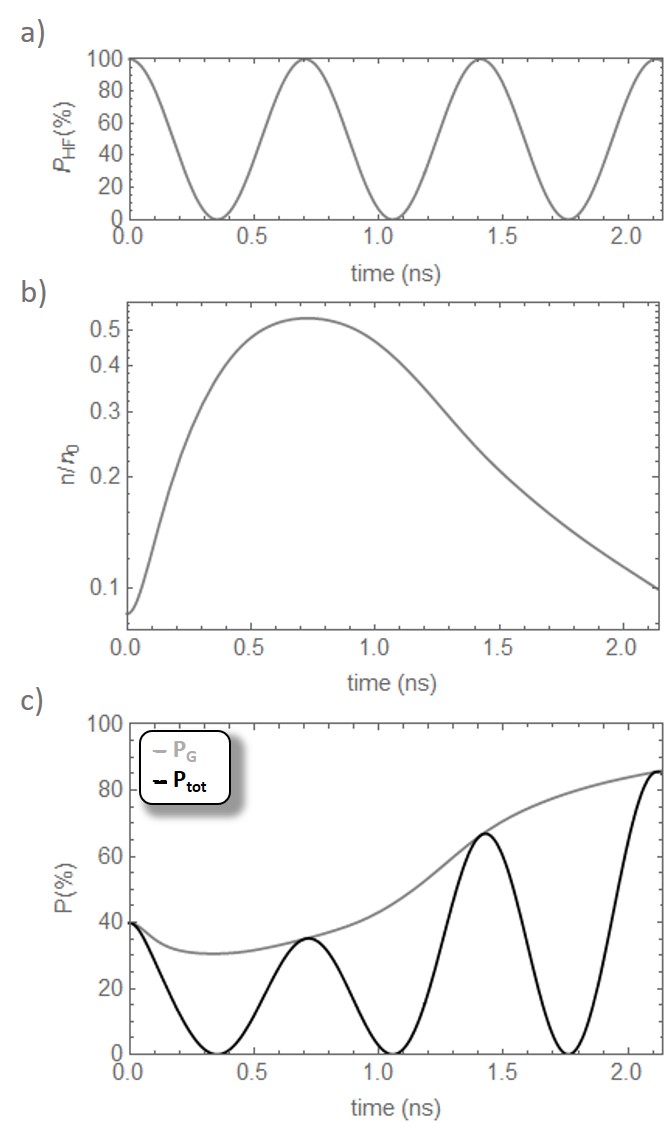}
	\caption{a) Polarization evolution of the hydrogen electron due to the hyperfine interaction b) Density of SPH atoms target atoms $n_T$  inside the hole, over the initial density of the reservoir $n_R$, as a function of time after firing DP2, in the absence of collisions. c) $P_G$ (dashed line) and total polarization (solid line) of the hydrogen atoms inside the hole as a function of time after firing the DP2, again, in the absence of collisions.}
	\label{FigureResults}
\end{figure}

Following dissociation, fast moving SPH atoms move towards all directions, with a large number of them ending up in the volume of the hole. The polarization of these atoms depends on their recoil angle: ns laser dissociation of HCl results in narrow velocity distributions for the atomic fragments \cite{HBrANDHCl}. For dissociation at $\lambda$ = 213 nm, the velocity distribution as a function of angle is $I(\theta) = N \left(1 + \frac{1}{2} P_2(\cos\theta)\right)$, with $P_2(x)$ being the second Legendre polynomials and $N$ a normalization factor. The polarization of the SPH atoms along the z direction depends on $\theta$ as P($\theta$) = $\cos ^2\theta$. In both cases, the angle $\theta$ is the apex angle used in spherical coordinates (shown in Fig.\ref{FigureSetUp}b), i.e.~the dissociation laser beams propagate parallel to the z axis with $\theta$ = 0. Note that, following dissociation, the polarization of the SPH oscillates due to coupling with the nuclear spin (1/2 for the proton), via the hyperfine interaction \cite{Qbeats}, as shown in Fig.\ref{FigureResults}a.

In Fig.\ref{FigureResults}b, we see the ratio of the target density $n_T$ of the SPH entering the hole over the initial density of HCl molecules in the reservoir $n_R$, as a function of time after firing DP2, in the absence of collisions. The density of the SPH atoms rises fast, reaching close to half the initial density of the HCl molecules in the molecular beam at $t\approx$ 0.5 ns, and afterwards is gradually reduced. 

The SPH polarization depends on (a) emission angle and (b) time, given by $P_{HF}(t)$ (shown in Fig.2a). We refer to the emission-angle polarization as \textit{geometric} polarization ($P_G$), and the total (observable) polarization as $P_{tot}$ = $P_G$$\times$$P_{HF}(t)$.

In Fig.\ref{FigureResults}c, we show the values for $P_G$ (dashed line) and $P_{tot}$ (solid line) as a function of time. We see that the $P_G$ curve starts around 40\% (the free-space value), and increases to $\approx$80\% at t $\approx$ 2 ns. This polarization increase is to be expected: the hole is shaped as a prolate ellipsoid with its large axis parallel to the laser propagation. SPH atoms recoiling at large angles with respect to the laser propagation axis have low polarization (P($\theta \approx \pi/2$)$\approx$ 0), and exit the hole very quickly. In contrast, highly polarized SPH atoms with small recoil velocities stay in the hole longer, since they transverse a much larger distance to exit\cite{FrontiersIsolated}. 

In Fig.\ref{FigureResults}c we see that the electron polarization is maximized at t = 0.7 ns, i.e.~the hyperfine period of the electron-proton system and all integer multiples of this time. The overall electron polarization at t = 1.42 ns is close of 70\% with a loss of a factor of $\approx$ 1/5 in density (shown in Fig.\ref{FigureResults}b), while at t = 2.12 ns the polarization surpasses 80\%, with an almost double corresponding reduction in density. Note that the solid curve reaches zero at t = 0.35 ns (half the hyperfine period for the electron-proton system) and integer multiples of this time. At this point all polarization is transferred to the proton nuclear spin, offering a target for laser acceleration of  polarized protons.

The maximum density in which such a pre-polarized target can be prepared is limited by elastic collisions. The H-H polarized elastic collision cross section is calculated to be around $\sigma_{H-H} \approx$130 a.u.\cite{DCSdatabase,HHelastic1,fox1967elasticH}, while similar value for the cross section $\sigma_{H-Cl}\approx$ 160 a.u., related to the elastic H-Cl is expected (see Supplemental Material). Note that, depolarizing H-Cl collisions are negligible in comparison, as the related cross section has been measured to be close to 2.8 a.u. \cite{spiliotis2021depolarization}. However, even non-depolarizing collisions, change the SPH recoil trajectories and ultimately break the correlation between the recoil angle and polarization. 
We can estimate the number of elastic collisions by considering the collisions mean-free-path $l_{mfp}$ and the mean-free-time between collisions $t_{fct}$. 

\begin{figure}[h!]
	\centering
	\includegraphics[width=0.45\textwidth]{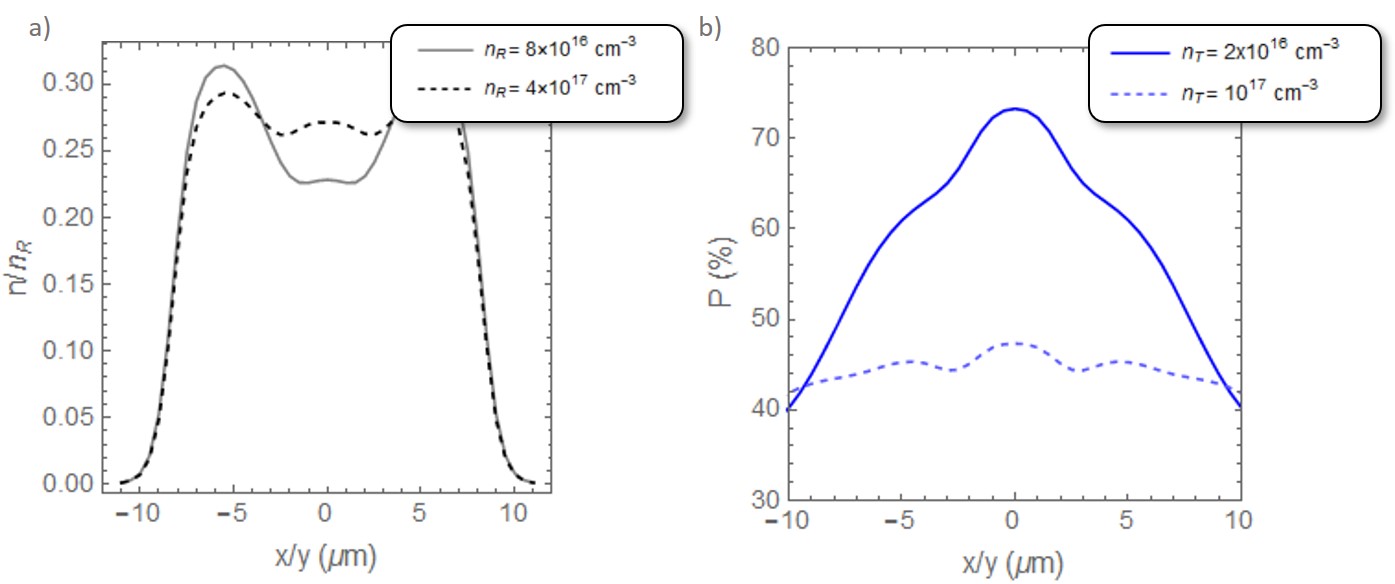}
	\caption{\label{fig:pic} a) Density distribution of the target SPH atoms $n_T$ at t = 1.42 ns, for $n_R$ = 4$\times$10$^{17}$ cm$^{-3}$(solid gray line) and  $n_R$ = 8$\times$10$^{16}$ cm$^{-3}$(dashed black line). b) Polarization distribution of the target SPH atoms at t = 1.42 ns, at $n_T$ = 10$^{17}$ cm$^{-3}$(solid blue line) and  $n_T$ = 2$\times$10$^{16}$ cm$^{-3}$(dashed blue line).}
	\label{FigCollisions}
\end{figure}

In Fig.\ref{FigCollisions}a, we show with the solid gray line, the target density $n_T$ at $t_{acc}$ = 1.42 ns, when the reservoir density (i.e. the initial density of the molecular beam) has been chosen to be $n_R$ = 8$\times$10$^{16}$ cm$^{-3}$. This density corresponds to $l_{mfp}$ $\approx$ 17 $\mu$m and $t_{fct}$ $\approx$ 1 ns. Looking at Fig.\ref{FigureResults}b, we see that by this time, the maximum density inside the hole has been reached, and it is starting to decline. This means that the by this time, most of the atoms will have entered the area of the hole, which lies in a lower density, and therefore they will experience almost no collisions. The evolution of the density inside the hole will mostly resemble the one shown in Fig.\ref{FigureResults}b where no collisions have been taken into account. Similarly, the polarization, shown in Fig.\ref{FigCollisions}b with the blue solid line, will have an evolution similar to what is shown in Fig.\ref{FigureResults}c, as only few atoms in the borders of the hole will have been affected by collisions. 

If we assume however a higher density, for example equal to $n_R$ = 4$\times$10$^{17}$ cm$^{-3}$, then $l_{mfp}$ $\approx$ 3.5 $\mu$m and $t_{fct}$ $\approx$ 0.2 ns. At this time, most SPH atoms will still be outside the volume of the hole, and therefore collisions have to be taken into account. Collisions randomize the emission direction and while they do not significantly change the density fraction of the atoms ending up in the hole at $t_{acc}$ = 1.42 ns, (shown in Fig.\ref{FigCollisions}a with a dashed black line), the polarization has been almost totally randomized and reduced to near it's free-space value of 40\%, as we see in Fig.\ref{FigCollisions}b with the dashed blue line.

\begin{figure}[htbp!]
	\centering
	\includegraphics*[width=0.45\textwidth]{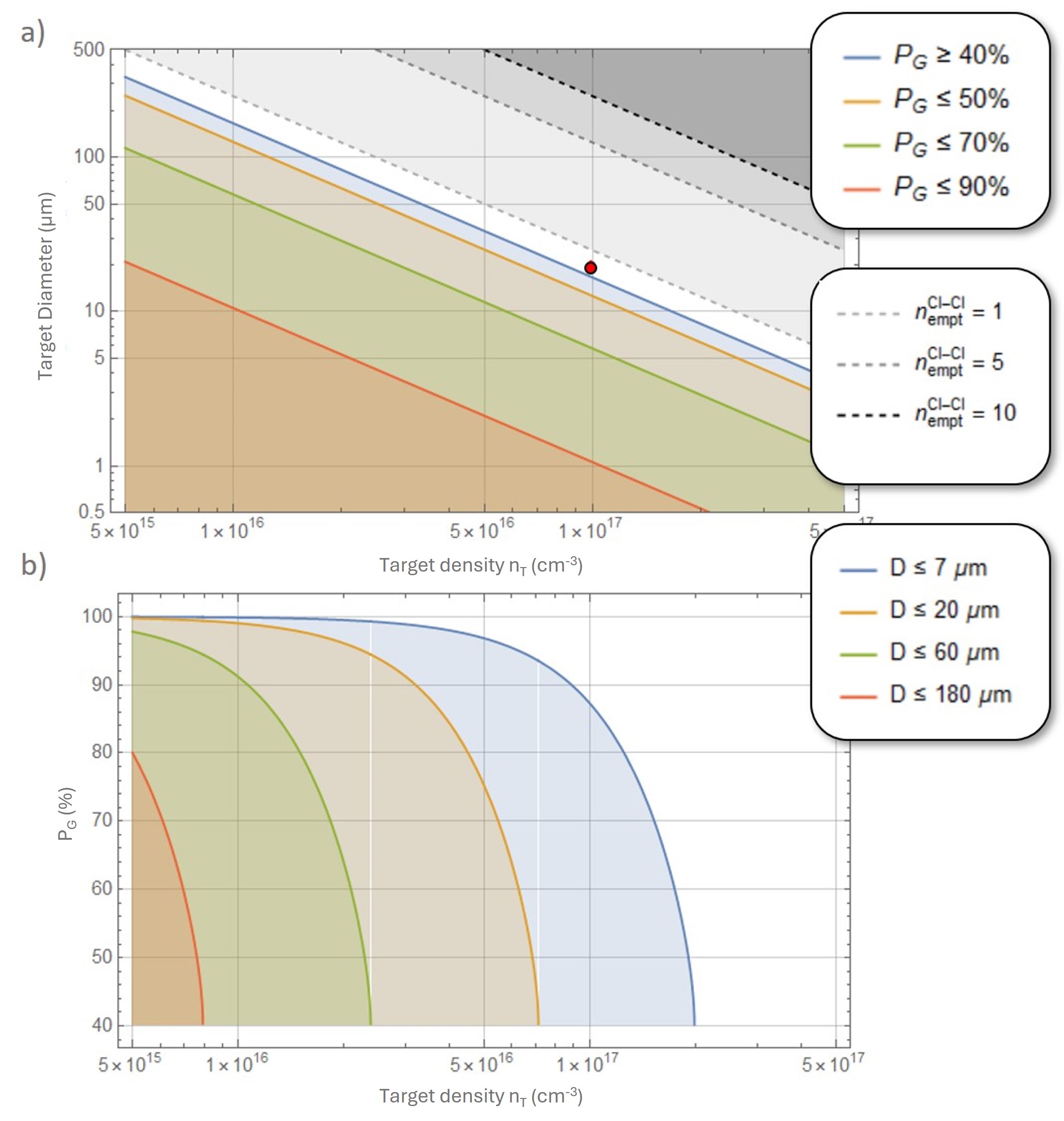}
	\caption{ a) Solid lines: Maximum SPH pre-polarized target diameter as a function of the target density, for various degrees of polarization. Dashed lines: Average number of Cl-Cl collisions $n_{empt}^{Cl-Cl}$ during the first step of creating the hole using DP1. The red dot shows the conditions in which PIC simulations were performed. b) Maximum polarization as a function of density for various choices for the target diameter D.    }
	\label{DiametersAtDensities}
\end{figure}

The target density and polarization depend both on the size and the angular dependence of the elastic collision differential cross section (DCS) \cite{fox1967elasticH}. We have found that for an isotropic DCS, a useful rule of thumb for estimating the target polarization as a function of its density and dimensions is to set the density so that $t_{acc}\approx$ 1.5$\times t_{cft}$, an arrangement that keeps collisional depolarization limited to the outer parts of the target. The target diameter, density and polarization regimes resulting from such a criterion are shown in Fig.\ref{DiametersAtDensities} using the solid lines. The maximum target diameter at a given density and polarization is approximately inversely proportional to the total elastic collision cross section. When the target polarization is chosen to be close to the thermal value os 40\%, the criterion  $t_{acc}\approx$1.5$t_{cft}$ can be relaxed even further to allow a few consecutive collision events, since for a target polarization of 40\%, polarization no longer limits the target size; however, new limitations arise from collisions during the first step, that of creating the hole using DP1. In this step, Cl-Cl elastic collisions can delay the emptying process by forcing the Cl atoms into a random walk, and thus, the process of creating the hole is expected to gradually become more difficult as the number of average collisions exceeds a few (let's say 5) collisions. These limitations are visualized by the gray dashed lines which mark the onset of one, five and ten average Cl-Cl collisions during the step of emptying the hole using DP1.

To investigate wakefield acceleration of polarized electrons with such targets, we conduct two-dimensional particle-in-cell (PIC) simulations with the code \textsc{vlpl} \cite{Pukhov1999, Pukhov2016}.
The simulations utilize a grid resolution of $h_x = 0.02 \lambda, h_y = 0.05 \lambda$. As we utilize the rhombi-in-plane Maxwell solver \cite{Pukhov2020}, the time step is chosen as $\Delta t = h_x / c$. The wavelength of the driving laser, which is the normalization constant for our simulations is chosen as $\lambda = 1.6$ {\textmu}m. Wakefields driven by CO$_2$ laser pulses or pulses in the mid-infrared range have been the subject of several theoretical studies like \cite{Brunetti2022, Papp2018} showing the self-trapping at lower plasma densities can be achieved, making use of the fact that the threshold for self-trapping in wakefields depends on the critical density, which is lower for larger wavelengths.

We model our target as a slab, consisting of Gaussian-shaped HCl walls with a density of $3.5 \times 10^{17}$ cm$^{-3}$ and a central channel containing the spin-polarized Hydrogen (and electrons), at a density of $10^{17}$ cm$^{-3}$ and a channel width of 20 {\textmu}m.
For simulation purposes, we choose a target length of 240 {\textmu}m and consider the SPH as being pre-ionized (potential spin-dependent effects during ionization are discussed i.a. in \cite{Klaiber2014}). The electrons are pre-polarized in $z$-direction. 

We use a driving pulse (moving in $+z$-direction) with $a_0 = 6$, focal spot size of $3 \lambda$ and a duration of $6 \lambda / c$. As shown in Fig. \ref{fig:pic}, polarized electrons are guided and accelerated in the channel-structure induced by the laser pulse.
Due to targetry restrictions, the choice of laser parameters is rather limited:
while stronger pulses are of interest for self-trapping, higher intensity will lead to increased spin precession and a loss of polarized electrons to the HCl walls.
For the aforementioned laser and target parameters, we are able to accelerate 3.9 pC up to approximately 4 MeV over the target length  and up to 48\% of the initial target polarization is preserved. Considering an initial polarization around $45$\% for the isolated SPH target, we result to a final polarization for the accelerated electron of $\approx$22\%. 
These electrons can be injected into a second wakefield stage (similar to \cite{Steinke2016, Lindstroem2021}) in order to obtain higher energies. If the Lorentz factor becomes sufficiently large during the first stage, the precession of spins according to the T-BMT equation becomes negligible \cite{Thomas2020}.

The laser parameters and injection scheme could be further tuned for different target densities: 
wider holes of SPH are possible to generate, however, at lower densities, as shown in Fig.\ref{DiametersAtDensities}. One scheme previously proposed for polarized HCl targets has been that of colliding-pulse injection \cite{Bohlen2023, Gong2023}. In these theoretical considerations, however, a fully pre-polarized target with $10^{18}$ cm$^{-3}$ density and much larger dimensions was used for simulations.
Further optimization of this target for wakefield acceleration could consist of implementing density ramps, for example by intensity-shaping \cite{Shaping1} the reservoir laser beam (see Supplemental Material) or using ionization injection as an alternative injection method. Finally, THz acceleration might offer the possibility of efficient acceleration at lower target densities, due to the scaling of critical density with the acceleration pulse wavelength \cite{THzScaling,nanni2015terahertz}, and allow accessing regimes of higher target polarization (see Fig.\ref{DiametersAtDensities}).  


\begin{figure}[h!]
	\centering
	\includegraphics[width=0.45\textwidth]{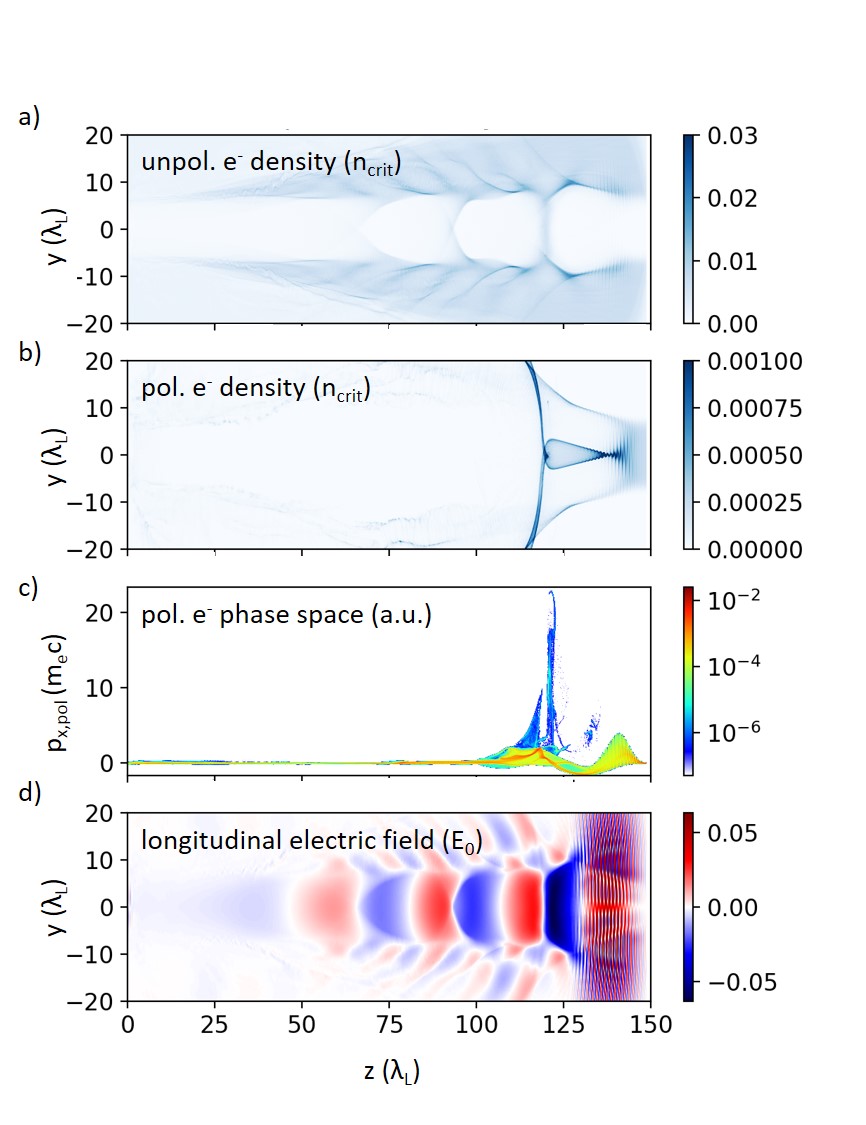}
	\caption{\label{fig:pic} PIC simulation results. Subplots (a), (b) show the unpolarized and polarized electron density, respectively. Plot (c) shows the phase space of only the pol. electrons, while (d) shows the transverse electric field. The colorbars are clipped for better visibility.}
	\label{PICfig}
\end{figure}

We have shown how, by combining the optical properties and the stereodynamics of the photodissociation process, one can prepare an isolated target of highly polarized SPH atoms, to be used in laser initiated electron acceleration experiments. We have  demonstrated how a sample suitable for a simple wakefield acceleration scheme can be prepared to allow MeV energies, which can be brought well into the GeV regime using a subsequent acceleration stage. Owing to the simplicity of the proposed method and the universality of the photodissociation dynamics, a large variety of similar pre-polarized targets can be designed to fit the needs of other acceleration schemes.   

 \subsection{Acknowledgments}
TPR acknowledges partial financial support  by the Hellenic Foundation for Research and Innovation (HFRI) and the General Secretariat for Research and Technology (GSRT), grant agreement No. HFRI-FM17-3709 (project NUPOL). The authors gratefully acknowledge the Gauss Centre for Supercomputing e.V. (\url{www.gauss-centre.eu}) for funding this project (spaf) by providing computing time through the John von Neumann Institute for Computing (NIC) on the GCS Supercomputer JUWELS at Jülich Supercomputing Centre (JSC).

\end{thebibliography}

	
\end{document}